\documentclass[12pt]{article}%
\usepackage{amsmath}
\usepackage{graphicx}%
\usepackage{amsfonts}%
\usepackage{amssymb}

\hoffset = - 8mm
\begin{document}

\title{ }

\begin{center}
\textbf{{\large RAMAN\ SCATTERING IN QUANTUM DISKS: ENHANCED EFFICIENCY OF THE
ELECTRON-PHONON INTERACTION DUE TO NON-ADIABATICITY}}

\bigskip

{\large S. N. Klimin}$^{1}${\large , V. M. Fomin}$^{1}${\large , J. T.
Devreese}$^{1}${\large , and J.-P. Leburton}$^{2}$

\bigskip

$^{1}${\large Theoretische Fysica van de Vaste Stof,}

{\large Universiteit Antwerpen (U.I.A.), B-2610 Antwerpen, Belgium}

\medskip

$^{2}${\large Department of Electrical and Computer Engineering}

{\large and Beckman Institute,}

{\large University of Illinois at Urbana-Champaign,}

{\large Urbana, IL 61801, USA}

\medskip

(November 15, 2001)

\vspace{1.5cm}

\textbf{ABSTRACT}
\end{center}

\begin{quote}
We treat resonant Raman scattering via the multiphonon exciton transitions in
cylindrical quantum dots with a parabolic confinement in the lateral direction
and with a finite rectangular interface-barrier confinement in the axial
direction. The optical phonons and the electron-phonon interaction are
considered within the multimode dielectric model. The model exploits both
electrostatic and mechanical boundary conditions for the relative ionic
displacement vector, as well as the phonon spatial dispersion in bulk. The
confined phonon modes in a quantum dot are hybrids of bulk-like and interface
vibrations. Raman spectra, calculated using the multimode dielectric model,
compare well with experimental data. Multiphonon Raman amplitudes are
calculated taking into account the effects of non-adiabaticity, which play a
crucial role in the optical spectra of quantum dots. Peak intensities of Raman
spectra are investigated as a function of the confinement frequency parameter
and of the height of a quantum dot. Non-adiabatic transitions are shown to
strongly increase the scattering probabilities and the relative multiphonon
scattering intensities with respect to the one-phonon intensity.
\end{quote}

\bigskip

\begin{center}
\textbf{I. INTRODUCTION}

\bigskip
\end{center}

Recent experiments on Raman scattering in self-assembled quantum dots
\cite{1,2,3,4,5,7} have demonstrated a rich structure of peaks provided by the
optical-phonon-assisted quantum transitions. In contrast to bulk, specific
phonon modes appear in quantum dots. In Ref. \cite{1}, an experimental
evidence is presented, that in pyramidal self-assembled InAs/GaAs quantum
dots, specific phonon modes exist whose frequencies differ from those both of
LO and of TO phonons. They are interpreted in terms of interface phonons
localized at the apexes of pyramids. Raman peaks assigned to interface phonons
are observed in Ref. \cite{7}. Raman spectra of disk-shape self-assembled
quantum dots, measured in Refs. \cite{2,3,4}, as shown in the present paper,
can be also interpreted in terms of phonon modes specific for quantum dots. In
Ref. \cite{5}, both one-phonon and two-phonon Raman bands are observed in
self-assembled InAs/GaAs quantum dots. This result confirms the conclusion of
Ref. \cite{Rapid} regarding an enhanced efficiency of the electron-phonon
interaction in quantum dots.

The dependence of this efficiency on sizes of a quantum dot is determined by
several factors. One of them is the following. When a finite-height potential
barrier is at the interface of a quantum dot, an enhanced separation of the
electron and hole charges can occur due to different masses, what enhances the
efficiency of the exciton-phonon interaction. The second factor of this
enhancement in quantum dots is the effect of non-adiabaticity \cite{LumQD}.
Various theoretical investigations lead to different conclusions on the
efficiency of the exciton-phonon interaction as a function of the quantum-dot
size, depending on a chosen model. In Ref. \cite{Klein}, a donor-like exciton
model for a spherical quantum dot has been considered, where a hole is treated
as a localized charge in the centre of a quantum dot. It is found within this
model, that the Huang-Rhys parameter does not depend on the quantum-dot
radius. On the contrary, in Ref. \cite{Marini}, for the same model, an
increase of the excitonic polaron effect with decreasing quantum dot size is
obtained using a variational method. The Raman cross section was studied for
weak confinement of an exciton in a CuBr quantum dot in Ref. \cite{Fedorov},
where it was shown that the Huang-Rhys parameter increased with the decrease
of the dot radius. The effects of non-adiabaticity of the exciton-phonon
system in spherical CdSe and PbS quantum dots in the strong-confinement regime
were shown in Ref. \cite{SSE} to lead to a significant rise of probabilities
of phonon-assisted transitions and to a substantial distinction of the Raman
and photoluminescence spectra from the Franck-Condon progression. The
enhancement of the exciton-phonon interaction makes possible an experimental
observation of multiphonon Raman scattering in quantum dots, what has been
confirmed by recent measurements of two-phonon Raman bands in self-organized
InAs/GaAs quantum dots \cite{5}.

In the present work, we investigate multiphonon resonant Raman scattering via
the exciton transitions in cylindrical quantum dots, which are a model for
self-assembled disk-shape quantum dots studied experimentally in Refs.
\cite{2,3,4}. Within our model, we assume a parabolic confinement in the
lateral direction and a finite rectangular interface-barrier confinement in
the axial direction. This model allows us to calculate exciton states in a
quantum dot by a variational method for a wide range of the
lateral-confinement frequency parameter from the weak-confinement regime to
the strong-confinement regime.

We interpret the experiments \cite{2,3,4} taking into account the effects of
non-adiabaticity, which have been shown in Ref. \cite{LumQD} to play a crucial
role in the optical spectra of quantum dots. We demonstrate, that different
additional channels of the phonon-assisted optical transitions open due to
non-adiabaticity of the exciton-phonon system in quantum dots.

\newpage

\begin{center}
\textbf{II. EXCITONS AND OPTICAL\ PHONONS IN\ QUANTUM\ DOTS}
\end{center}

\bigskip

We consider a cylindrical quantum dot of the height $h$, which occupies the
space $-h/2<z<h/2$ along the $z$-axis. The lateral confinement for electrons
and holes is ensured by a parabolic potential with the frequency parameter
$\Omega_{0}.$ A medium inside (outside) the quantum dot is indicated by the
index 1 (2).

The exciton-phonon Hamiltonian of the system under consideration has the form%
\begin{equation}
\hat{H}=\hat{H}_{ex}\left(  \mathbf{r}_{e},\mathbf{r}_{h}\right)  +\sum_{\nu
}\hbar\omega_{\nu}\hat{a}_{\nu}^{\dag}\hat{a}_{\nu}+\sum_{\nu}\left(
\beta_{\nu}\hat{a}_{\nu}+\beta_{\nu}^{\ast}\hat{a}_{\nu}^{\dag}\right)  ,
\label{Ham}%
\end{equation}
where $H_{ex}\left(  \mathbf{r}_{e},\mathbf{r}_{h}\right)  $ is the ``bare''
exciton Hamiltonian in the effective-mass approach, $\mathbf{r}_{e}$ and
$\mathbf{r}_{h}$ are the coordinates of an electron and of a hole,
respectively;%
\begin{align}
\hat{H}_{ex}\left(  \mathbf{r}_{e},\mathbf{r}_{h}\right)   &  =-\frac{\hbar
^{2}}{2}\frac{\partial}{\partial z_{e}}\left(  \frac{1}{m_{e}\left(  z\right)
}\frac{\partial}{\partial z_{e}}\right)  -\frac{\hbar^{2}}{2}\frac{\partial
}{\partial z_{h}}\left(  \frac{1}{m_{h}\left(  z\right)  }\frac{\partial
}{\partial z_{h}}\right) \nonumber\\
&  -\frac{\hbar^{2}}{2m_{e}\left(  z\right)  }\left(  \frac{\partial^{2}%
}{\partial x_{e}^{2}}+\frac{\partial^{2}}{\partial y_{e}^{2}}\right)
-\frac{\hbar^{2}}{2m_{h}\left(  z\right)  }\left(  \frac{\partial^{2}%
}{\partial x_{h}^{2}}+\frac{\partial^{2}}{\partial y_{h}^{2}}\right)
+U_{c}\left(  \mathbf{r}_{e},\mathbf{r}_{h}\right) \nonumber\\
&  +\frac{m_{e}\left(  z_{e}\right)  \Omega_{0}^{2}}{2}\mathbf{r}_{e\parallel
}^{2}+\frac{m_{h}\left(  z_{h}\right)  \Omega_{0}^{2}}{2}\mathbf{r}%
_{h\parallel}^{2}+V_{e}\left(  z_{e}\right)  +V_{h}\left(  z_{h}\right)
\label{Hex}%
\end{align}
with the Coulomb electron-hole interaction potential $U_{c}\left(
\mathbf{r}_{e},\mathbf{r}_{h}\right)  $ and with the interface barrier
potential:%
\begin{equation}
V_{e\left(  h\right)  }\left(  z\right)  =\left\{
\begin{array}
[c]{c}%
0,\quad\left|  z\right|  <h/2,\\
V_{e\left(  h\right)  }^{b},\quad\left|  z\right|  \geq h/2.
\end{array}
\right.  \label{BP}%
\end{equation}
In Eq. (\ref{Hex}), the notation $m_{e\left(  h\right)  }\left(  z\right)  $
means that we take into account a distinction between electron (hole)
effective masses in the quantum dot (medium 1) and in a host substance (medium
2):%
\begin{equation}
m_{e\left(  h\right)  }\left(  z\right)  =\left\{
\begin{array}
[c]{c}%
m_{e\left(  h\right)  1},\quad\left|  z\right|  <h/2,\\
m_{e\left(  h\right)  2},\quad\left|  z\right|  \geq h/2.
\end{array}
\right.  \label{mass}%
\end{equation}

In the present work, we consider the case when the height of the quantum dot
is much less than both the exciton Bohr radius and effective lateral radius of
the quantum dot $R_{0}\sim\sqrt{\hbar/\left(  2\mu_{1}\Omega_{0}\right)  },$
where $\mu_{1}$ is the reduced mass of an electron and of a hole in the
quantum dot. Eigenstates of the exciton Hamiltonian (\ref{Hex}), which are
used for the calculation of the Raman spectra, are found by a variational
method, assuming a strong-confinement regime along the $z$-axis. We choose the
variational wave functions in the form%
\begin{align}
\Psi_{N_{e}N_{h},n_{1}m_{1},n_{2}m_{2}}\left(  \mathbf{r}_{e},\mathbf{r}%
_{h}\right)   &  =\phi_{N_{e}}^{e}\left(  z_{e}\right)  \phi_{N_{h}}%
^{h}\left(  z_{h}\right) \nonumber\\
&  \times\Phi_{n_{1}m_{1}}\left(  R\right)  \frac{1}{\sqrt{2\pi}}%
e^{im_{1}\varphi_{1}}\psi_{n_{2}m_{2}}\left(  \rho\right)  \frac{1}{\sqrt
{2\pi}}e^{im_{2}\varphi_{2}}, \label{WaveFun}%
\end{align}
where $\phi_{N_{e\left(  h\right)  }}^{e\left(  h\right)  }\left(  z\right)  $
are the eigenfunctions of the Hamiltonian of the electron (hole) motion along
the $z$-axis:%
\begin{equation}
\hat{H}_{e\left(  h\right)  }^{\perp}\left(  z\right)  \equiv-\frac{\hbar^{2}%
}{2}\frac{\partial}{\partial z}\left(  \frac{1}{m_{e\left(  h\right)  }\left(
z\right)  }\frac{\partial}{\partial z}\right)  +V_{e\left(  h\right)  }\left(
z\right)  \label{Hperp}%
\end{equation}
with the current-conserving boundary conditions at the interface
\cite{Ando82}:%
\begin{align}
&  \left\{
\begin{array}
[c]{c}%
\phi_{N_{e\left(  h\right)  }}^{e\left(  h\right)  }\left(  \pm\left(
\frac{h}{2}-\delta\right)  \right)  =\phi_{N_{e\left(  h\right)  }}^{e\left(
h\right)  }\left(  \pm\left(  \frac{h}{2}+\delta\right)  \right)  ,\\
\left.  \frac{1}{m_{e\left(  h\right)  }\left(  z\right)  }\frac{\partial
}{\partial z}\phi_{N_{e\left(  h\right)  }}^{e\left(  h\right)  }\left(
z\right)  \right|  _{z=\pm\left(  \frac{h}{2}-\delta\right)  }=\left.
\frac{1}{m_{e\left(  h\right)  }\left(  z\right)  }\frac{\partial}{\partial
z}\phi_{N_{e\left(  h\right)  }}^{e\left(  h\right)  }\left(  z\right)
\right|  _{z=\pm\left(  \frac{h}{2}+\delta\right)  }%
\end{array}
\right. \label{BBC}\\
&  \left(  \delta\rightarrow+0\right)  .\nonumber
\end{align}
The functions $\Phi_{n_{1}m_{1}}\left(  R\right)  \frac{1}{\sqrt{2\pi}%
}e^{im_{1}\varphi_{1}}$ and $\psi_{n_{2}m_{2}}\left(  \rho\right)
\frac{1}{\sqrt{2\pi}}e^{im_{2}\varphi_{2}}$ describe, respectively, quantum
states of the ``in-plane'' center-of-mass and relative motions of an electron
and of a hole. $m_{1}$ and $m_{2}$ are quantum numbers of the angular
momentum, while $n_{1}$ and $n_{2}$ are ``radial'' quantum numbers. The wave
functions $\Phi_{n_{1}m_{1}}\left(  R\right)  $ and $\psi_{n_{2}m_{2}}\left(
\rho\right)  $ contain variational parameters.

In Eq. (\ref{Ham}), $\omega_{\nu}$ are frequencies of the phonon modes, which
are specific for the quantum dot and numbered by the index $\nu$. $\beta_{\nu
}=\gamma_{\nu}\left(  \mathbf{r}_{e}\right)  -\gamma_{\nu}\left(
\mathbf{r}_{h}\right)  $ is the exciton-phonon interaction amplitude. Within
the present investigation, optical phonons in the quantum dot are considered
using the \emph{multimode continuum model} \cite{PSS95,JLum}. In the case when
the sizes of the quantum dot substantially exceed the lattice constant
$a_{0},$ optical phonons both in the quantum dot and in the host medium are
described by the relative ionic displacement vector $\mathbf{u}\left(
\mathbf{r}\right)  =\mathbf{u}_{+}\left(  \mathbf{r}\right)  -\mathbf{u}%
_{-}\left(  \mathbf{r}\right)  ,$ where $\mathbf{u}_{\pm}\left(
\mathbf{r}\right)  $ are displacements of the positive and negative ions,
respectively. The dynamics of the dispersive optical phonons in the long-wave
region is determined by the generalized Born-Huang equation (cf.
\cite{Chamberlain,Trallero}). It has the following form in the Fourier
representation with respect to time:
\begin{equation}
\left(  \omega_{\mathrm{TO}}^{2}-\omega^{2}\right)  \mathbf{u}=\omega
_{\mathrm{TO}}\left(  \frac{\varepsilon_{0}-\varepsilon_{\infty}}{4\pi\rho
}\right)  ^{1/2}\mathbf{E}-\nabla\cdot\mathcal{S}\left(  \mathbf{u}\right)  ,
\label{BH}%
\end{equation}
where $\rho$\ is the reduced ionic mass density, $\mathbf{E}$\ is the
macroscopic electric field, $\omega_{\mathrm{T}}$ is the bulk TO phonon
frequency at the Brillouin zone centre, $\varepsilon_{\infty}$\ and
$\varepsilon_{0}$\ are the high-frequency and static dielectric constants,
respectively. The stress tensor
\begin{equation}
\left[  \mathcal{S}\left(  \mathbf{u}\right)  \right]  _{ij}\equiv\left(
v_{\mathrm{L}}^{2}-2v_{\mathrm{T}}^{2}\right)  \left(  \nabla\cdot
\mathbf{u}\right)  \delta_{ij}+v_{\mathrm{T}}^{2}\left(  \frac{\partial u_{i}%
}{\partial x_{j}}+\frac{\partial u_{j}}{\partial x_{i}}\right)  \label{tensor}%
\end{equation}
corresponds the LO and TO bulk phonon dispersion characterized by parameters
$v_{\mathrm{L}}$ and $v_{\mathrm{T}}$, respectively:%
\begin{equation}
\left\{
\begin{array}
[c]{c}%
\omega_{\mathrm{LO}}^{2}\left(  q\right)  =\omega_{\mathrm{LO}}^{2}\left(
0\right)  -v_{\mathrm{L}}^{2}q^{2},\\
\omega_{\mathrm{TO}}^{2}\left(  q\right)  =\omega_{\mathrm{TO}}^{2}\left(
0\right)  -v_{\mathrm{T}}^{2}q^{2},
\end{array}
\right.  \label{Bulk}%
\end{equation}
where $q$ is the modulus of the phonon wave vector. Eigenfrequencies and basis
vectors of phonon modes are found by the joint solution of the equation
(\ref{BH}) and of the static Maxwell equation for the displacement
$\mathbf{D=}\varepsilon_{\infty}\mathbf{E}+\omega_{\mathrm{TO}}\sqrt
{4\pi\left(  \varepsilon_{0}-\varepsilon_{\infty}\right)  \rho}\mathbf{u}$,
\begin{equation}
\mathrm{div\,}\mathbf{D}=0. \label{Maxwell}%
\end{equation}

For the quantum dot structure, Eqs. (\ref{BH}) and (\ref{Maxwell}) should be
obeyed for each medium separately. The solution of this set of equations needs
boundary conditions. They can be subdivided into electrostatic [for
$\mathbf{E}\left(  \mathbf{r}\right)  $ and $\mathbf{D}\left(  \mathbf{r}%
\right)  $] and mechanical [for $\mathbf{u}\left(  \mathbf{r}\right)  $]
boundary conditions.

Eigenvalues $\omega^{2}$ of the resulting boundary problem should be real.
This condition leads to the equation, which matches the values of the
functional $G\left(  \mathbf{u},\mathbf{u}^{\prime}\right)  \equiv
\rho\mathbf{u}^{\prime}\cdot\mathbf{F}\left(  \mathbf{u}\right)  $ on both
sides from the boundary between media 1 and 2,
\begin{equation}
\left.  \left(  G_{1}-G_{2}\right)  \right|  _{\mathrm{boundary}}=0.
\label{Hermit}%
\end{equation}
Here $\mathbf{F}\left(  \mathbf{u}\right)  \equiv\mathcal{S}\left(
\mathbf{u}\right)  \cdot\mathbf{n}$ is the force flux across the boundary,
$\mathbf{n}$ is the unit vector normal with respect to the boundary.

For structures where frequencies of optical phonons in adjacent media
substantially differ from each other, optical vibrations cannot propagate from
one medium to another. Hence, in this case we can set $\left.  G\right|
_{\mathrm{boundary}}=0$. This approximation is equivalent to model boundary
conditions proposed in Refs. \cite{Chamberlain,Trallero,Constantinou}. The
electrostatic interaction between lattice cells is modelled by the interaction
between dipoles, which is strongly anisotropic. Taking into account this
anisotropy, the following boundary conditions \cite{Constantinou} are
applicable within this model:
\begin{equation}
\left.  u_{\perp}\right|  _{\mathrm{boundary}}=0,\quad\left.  \mathbf{F}%
_{\parallel}\right|  _{\mathrm{boundary}}=0, \label{mbc}%
\end{equation}
where subscripts $\perp$ and $\parallel$ denote normal and tangential (with
respect to the boundary) components of a vector. For long-wavelength
vibrations, the second equation of (\ref{mbc}) transforms into $\left.
\partial\mathbf{u}_{\parallel}/\partial\xi_{n}\right|  _{\mathrm{boundary}}%
=0$, where $\xi_{n}$ is the coordinate directed along $\mathbf{n}$.

Eqs. (\ref{BH}) and (\ref{Maxwell}) are solved by substituting into them
$\mathbf{u}\left(  \mathbf{r}\right)  $ as a sum over basis vectors, which
satisfy the boundary conditions (\ref{mbc}). As distinct from the dielectric
continuum model (see, e. g. Ref. \cite{Lassnig}), we find that LO phonon modes
are not bulk-like or interface but have a hybrid character.

From Eqs. (\ref{BH}) and (\ref{Maxwell}) with (\ref{mbc}), dispersion
equations are derived for the eigenfrequencies of ``symmetric'' ($s$) and
``antisymmetric'' ($a$) modes, whose electrostatic potentials are,
respectively, symmetric and antisymmetric with respect to the reflection in
the $xy$-plane:%

\[
\varepsilon_{1s}\left(  q_{\parallel},\omega\right)  \tanh\left(  \frac{\zeta
}{2}\right)  +\varepsilon_{2}\left(  \omega\right)  =0\quad
\mbox {for
symmetric modes,}
\]%

\begin{equation}
\varepsilon_{1a}\left(  q_{\parallel},\omega\right)  \coth\left(  \frac{\zeta
}{2}\right)  +\varepsilon_{2}\left(  \omega\right)  =0\quad
\mbox {for
antisymmetric modes,}  \label{DispEq}%
\end{equation}
where $\zeta=q_{\parallel}h$, $q_{\parallel}$ is the modulus of the
``in-plane'' phonon wave vector $\mathbf{q}_{\parallel}$, and $\varepsilon
_{2}\left(  \omega\right)  $ is the dielectric function of a host medium. The
functions $\varepsilon_{1s\left(  a\right)  }\left(  q_{\parallel}%
,\omega\right)  $ have the form%

\begin{equation}
\varepsilon_{1j}\left(  q_{\parallel},\omega\right)  =\varepsilon_{1}\left(
\infty\right)  \left[  1-\sum_{n=0}^{n_{\max}}\chi_{j,n}\left(  q_{\parallel
}\right)  \frac{\omega_{1,\mathrm{LO}}^{2}-\omega_{1,TO}^{2}}{\omega
_{1,\mathrm{LO}}^{2}-v_{1,\mathrm{LO}}^{2}Q_{j,n}^{2}-\omega^{2}}\right]
^{-1} \label{Summa}%
\end{equation}
with%

\begin{equation}
Q_{j,n}^{2}\equiv\left\{
\begin{array}
[c]{ll}%
q_{\parallel}^{2}+\left(  \frac{2n\pi}{h}\right)  ^{2}, & j=s\\
q_{\parallel}^{2}+\left[  \frac{\left(  2n+1\right)  \pi}{h}\right]  ^{2}, &
j=a
\end{array}
\right.  .
\end{equation}
The coefficients $\chi_{j,n}\left(  q_{\parallel}\right)  $ are determined by%

\begin{equation}
\chi_{s,n}\left(  q_{\parallel}\right)  =\frac{2\left(  2-\delta_{n.0}\right)
q_{\parallel}}{hQ_{s,n}^{2}}\tanh\left(  \frac{\zeta}{2}\right)  ,\qquad
\chi_{a,n}\left(  q_{\parallel}\right)  =\frac{4q_{\parallel}}{hQ_{a,n}^{2}%
}\coth\left(  \frac{\zeta}{2}\right)  .
\end{equation}
The summation over $n$ in Eq. (\ref{Summa}) is limited by the value $n_{0}\sim
h/\left(  2a_{0}\right)  $. This limitation expresses the fact that the
wavelength of an optical phonon cannot be smaller than twice the lattice
constant $a_{0}$.

Using a formal analogy of the dispersion equations (\ref{DispEq}) with those
for interface phonon modes of the dielectric continuum model (see, e. g., Ref.
\cite{Lassnig}), we can interpret $\varepsilon_{1j}\left(  q_{\parallel
},\omega\right)  $ as the effective dielectric function of the quantum dot. As
distinct from the dielectric function in bulk, there is a set of dielectric
functions in the quantum dot, which describe polarization response due to
symmetric ($j=s$) and antisymmetric ($j=a$) optical-phonon modes,
respectively. It is worth noting, that the functions $\varepsilon_{1j}\left(
q_{\parallel},\omega\right)  $ are dispersive, since they depend both on
$\omega$ and on the ``in-plane'' wave vector $\mathbf{q}_{\parallel}.$

\bigskip

\begin{center}
\textbf{III. MULTIPHONON RAMAN\ INTENSITIES}

\bigskip
\end{center}

Within the context of the long-wavelength approximation, the interaction of an
electron with an electromagnetic field is described by the operator $\hat
{V}\left(  t\right)  =\hat{V}_{I}e^{-i\Omega_{I}t}+\hat{V}_{S}^{\dagger
}e^{i\Omega_{S}t}$, where the terms $\hat{V}_{I}$ and $\hat{V}_{S}^{\dagger}$
correspond, respectively, to the absorption of a photon with the frequency
$\Omega_{I}$ (incoming light) and to the emission of a photon with the
frequency $\Omega_{S}$ (scattered light). The interaction amplitude $\hat
{V}_{I(S)}$ is proportional to the projection of the electron dipole momentum
operator $\mathbf{\hat{d}}$ on the polarization vector $\mathbf{e}^{I(S)}$ of
the relevant wave: $\hat{d}^{I(S)}=\mathbf{e}^{I(S)}\cdot\mathbf{\hat{d}}$.
From the second-order perturbation theory, the transition probability between
an initial $\left|  i\right\rangle $ state and a final $\left|  f\right\rangle
$ state is
\begin{equation}
w_{i\rightarrow f}=\frac{2\pi}{\hbar^{4}}\left|  \sum\limits_{m}%
\frac{\left\langle f\left|  \hat{V}_{S}^{\dagger}\right|  m\right\rangle
\left\langle m\left|  \hat{V}_{I}\right|  i\right\rangle }{\omega_{mi}%
-\Omega_{I}+i\varepsilon}\right|  ^{2}\delta\left(  \omega_{fi}-\Omega
_{I}+\Omega_{S}\right)  ;\quad\varepsilon\rightarrow+0. \label{Probab}%
\end{equation}
Here, $\omega_{mi}$ and $\omega_{fi}$ are transition frequencies. For
phonon-assisted Raman scattering in the quantum dot, $\left|  i\right\rangle
$, $\left|  f\right\rangle $, and $\left|  m\right\rangle $ are quantum states
of the exciton-phonon system. Both the initial and the final states contain no
charge carriers (electrons or holes), so that these states are described by a
direct product of a wave function of free phonons with that of the exciton
vacuum. Intermediate states $\left|  m\right\rangle $ are eigenstates of the
total Hamiltonian (\ref{Ham}) of the exciton-phonon system. For definite
polarizations of the incoming and the scattered light, the scattering
probability is obtained by averaging Eq. (\ref{Probab}) over the initial
states and by summing over the final ones. Because electron-phonon coupling
constants in such semiconductors as InAs, GaAs, AlAs (which are often used
materials for self-assembled quantum dots) are small, the $K$-phonon
scattering intensity, corresponding to a definite combinatorial frequency
$\sum\limits_{j=1}^{K}\omega_{\nu_{j}}$, can be analyzed to the lowest
($K$-th) order in the electron-phonon coupling constants $\alpha_{1}$ and
$\alpha_{2}$. The scattering intensity is then expressed through a squared
modulus of the scattering amplitude:
\begin{equation}
F_{K}^{\left(  \pm\right)  }\left(  \nu_{1},...,\nu_{K}\right)  \,=\,\sum
_{\mu_{0}...\mu_{K}}\frac{d_{\mu_{0}}^{I}\left(  d_{\mu_{K}}^{S}\right)
^{\ast}}{\tilde{\omega}_{\mu_{0}}-\Omega_{I}+i\tilde{\Gamma}_{\mu_{0}}}%
\prod_{j=1}^{K}\frac{\left\langle \mu_{j}\left|  \hat{\beta}_{\nu_{j}}\right|
\mu_{j-1}\right\rangle }{\tilde{\omega}_{\mu_{j}}-\Omega_{I}\pm\sum
\limits_{k=1}^{j}\left(  \omega_{\nu_{k}}\pm i\Gamma_{\nu_{k}}\right)
+i\tilde{\Gamma}_{\mu_{j}}}. \label{F}%
\end{equation}
Here $\tilde{\omega}_{\mu}$ is the frequency and $d_{\mu}^{I(S)}\equiv
\langle\mu|{\hat{d}}^{I(S)}|0\rangle$ is the dipole matrix element of a
transition from the exciton vacuum to the eigenstate $|\mu\rangle$ of the
Hamiltonian $\hat{H}_{ex}$, $\tilde{\Gamma}_{\mu}$ is the inverse lifetime of
an exciton in the state $\left|  \mu\right\rangle $, $\Gamma_{\nu}$ is the
inverse lifetime of a phonon of the $\nu$-th mode.

\bigskip

\begin{center}
\textbf{IV. RESULTS AND DISCUSSION}

\bigskip
\end{center}

In Figs. 1 to 3, we have plotted one- and two-phonon Raman spectra for
cylindrical GaAs/AlAs quantum dots with different values of the dot height $h$
and of the lateral-confinement frequency $\Omega_{0}.$ In Fig. 1, the values
of these parameters are chosen the same as typical values given in Ref.
\cite{3}: $h=5$ nm, $\hbar\Omega_{0}=3$ meV. The material parameters of
semiconductors are taken from Ref. \cite{Landolt}. In particular, LO and TO
phonon frequencies at the Brillouin zone centre are: $\omega_{\mathrm{LO}%
}^{\left(  \mathrm{GaAs}\right)  }=292.37$ cm$^{-1}$, $\omega_{\mathrm{TO}%
}^{\left(  \mathrm{GaAs}\right)  }=268.5$ cm$^{-1}$, $\omega_{\mathrm{LO}%
}^{\left(  \mathrm{AlAs}\right)  }=403.7$ cm$^{-1}$, $\omega_{\mathrm{TO}%
}^{\left(  \mathrm{AlAs}\right)  }=361.7$ cm$^{-1}$. These frequencies are
indicated in the figures by arrows. Raman intensities are plotted as a
function of the Stokes frequency shift $\left(  \Omega_{I}-\Omega_{S}\right)
$.

In order to analyze a role of non-adiabaticity, Raman spectra have been
calculated by two methods: (i) within the non-adiabatic approach, using the
Raman amplitude (\ref{F}), where transitions through all intermediate states
are taken into account, and (ii) in the adiabatic approximation (see Ref.
\cite{Klein}), when non-diagonal matrix elements of the exciton-phonon
interaction in Eq. (\ref{F}) are neglected. The results of non-adiabatic and
adiabatic approaches are shown in Figs. 1 to 3 by solid and dashed lines, respectively.

In the one-phonon spectra, we can clearly distinguish between two groups of
peaks, whose frequencies lie in the regions of optical-phonon frequencies of
GaAs and of AlAs, respectively. Consequently, these peaks can be assigned to
the so-called ``GaAs-like'' and ``AlAs-like'' phonon modes. The structure of
the GaAs-like group of peaks reflects the spectrum of hybrid phonon modes. The
hybridization of the AlAs-like phonons appears to be negligibly small (as
distinct from the GaAs-like phonons), so that they are subdivided into the
half-space and interface AlAs-like phonons. The interaction of an exciton with
the half-space AlAs-like phonons is comparatively weak, because it is
determined only by ``tails'' of the electron and hole wave functions outside
the quantum dot.

Owing to appreciable optical-phonon dispersion in GaAs, hybrid GaAs-like modes
of the quantum dot occupy a wider frequency range than the gap between LO and
TO phonon frequencies at the Brillouin zone centre. It appears, that a hybrid
mode always exists, which frequency is close to the TO-phonon frequency of
GaAs. This mode has a larger dispersion than that for other hybrid GaAs-like
modes and provides a Raman peak of a considerable relative intensity. In the
limit of large $h,$ when the dielectric continuum model of optical phonons is
applicable, the aforesaid mode turns into the GaAs-like interface mode.
Consequently, within our model, this mode is formed mainly by interface
vibrations and can be referred to as the ``interface-like'' mode. It is worth
noting, that the experimentally observed Raman spectra of disk-shape
self-assembled Ga$_{0.8}$In$_{0.2}$As/Ga$_{0.5}$Al$_{0.5}$As quantum dots
\cite{3,4} (see the inset in Fig. 1) reveal a structure similar to that of the
``GaAs-band'' calculated within the present model. Namely, there is a
comparatively narrow intense peak near $\omega_{\mathrm{LO}}^{\left(
\mathrm{GaAs}\right)  }$ and a wider peak between $\omega_{\mathrm{TO}%
}^{\left(  \mathrm{GaAs}\right)  }$ and $\omega_{\mathrm{LO}}^{\left(
\mathrm{GaAs}\right)  }.$ Within our model, the former peak corresponds to the
most-long-wavelength hybrid mode (whose amplitude is mainly confined to the
quantum dot), while the latter is assigned to the aforesaid interface-like mode.

In Fig. 2, we show Raman spectra for the quantum dot of the same height as in
Fig. 1 but with a stronger lateral confinement: $\hbar\Omega_{0}=10$ meV. When
strengthening the lateral confinement, the relative contributions of the
interface-like phonons into Raman spectra diminish. In Fig. 3, we consider the
quantum dot with a smaller height than that in Fig. 1 ($h=2$ nm) and with the
same value of the lateral-confinement frequency $\hbar\Omega_{0}=3$ meV. In
this case, the relative intensity of peaks, assigned to the AlAs-like
interface modes, substantially increases. We can conclude, that the relative
contribution of the interface-like phonons to Raman spectra is an increasing
function of the aspect ratio $R/h.$

The two-phonon Raman band consists of peaks related to different combinatorial
optical-phonon frequencies. Note that peaks assigned to combinations of two
different phonon modes can be higher than those related to double frequencies
of each phonon mode (see, for example, the middle two-phonon band in Fig. 3).
This fact is explained as follows. The light scattering by two different
phonon modes can be realized through two channels, which differ from each
other by the permutation between the phonon modes. Hence, the factor 2 appears
for the corresponding scattering intensity when compared with that for the
subsequent scattering by one and the same phonon mode.

We see from Figs. 1 to 3 (where all spectra are plotted using one and the same
scale), that the effects of non-adiabaticity considerably enhance both the
absolute values of Raman peak intensities and the relative intensities of the
two-phonon peaks with respect to those of the one-phonon peaks. In Fig. 4, we
have plotted the parameter $S\equiv2I_{2}/I_{1},$ (where $I_{1}$ and $I_{2}$
are integral intensities of the one-phonon and two-phonon Raman bands,
respectively) as a function of the lateral-confinement frequency for different
values of the height of the quantum dot. The parameter $S$ is a measure of the
efficiency of the exciton-phonon interaction in the quantum dot. Within the
adiabatic approximation, $S$ coincides with the Huang-Rhys parameter%
\begin{equation}
S^{\left(  \mathrm{A}\right)  }=\sum_{\nu}\frac{\left|  \left\langle \mu
_{0}\left|  \hat{\beta}_{\nu}\right|  \mu_{0}\right\rangle \right|  ^{2}%
}{\left(  \hbar\omega_{\nu}\right)  ^{2}}, \label{HRP}%
\end{equation}
where $\left|  \mu_{0}\right\rangle $ is the ground state of an exciton in the
quantum dot. One can see, that $S,$ calculated accounting for non-adiabatic
transitions, strongly increases with strengthening confinement, and achieves
the value $S\sim0.05$ for $\hbar\Omega_{0}=10$ meV and $h=2$ nm. The ratio of
the ``non-adiabatic'' and ``adiabatic'' parameters $S^{\left(  \mathrm{NA}%
\right)  }/S^{\left(  \mathrm{A}\right)  },$ as seen from the inset in Fig. 4,
diminishes with increasing the confinement frequency. This can be explained by
an increasing energy difference between various quantum levels of an exciton
with increasing $\Omega_{0},$ what lowers the relative probabilities of
non-adiabatic transitions. Nevertheless, in the range of $\Omega_{0}$ studied,
$S^{\left(  \mathrm{NA}\right)  }>S^{\left(  \mathrm{A}\right)  }$.

Comparing Figs. 2 and 3 between each other, we see that with strengthening
confinement there occurs a substantial rise of scattering intensities,
especially of two-phonon ones. This result suggests that an experimental
observation of multiphonon Raman bands is facilitated in small-size
self-assembled quantum dots.

\bigskip

\begin{center}
\textbf{V. CONCLUSIONS}

\bigskip
\end{center}

We have treated multiphonon Raman spectra in cylindrical quantum dots, taking
into account non-adiabatic effects and specific optical-phonon modes, which
are considered within the multimode dielectric model \cite{PSS95,JLum}.

The relative contributions of different phonon modes into Raman spectra depend
both on the quantum-dot volume and on the aspect ratio of the radius to the
height of the quantum dot. The relative contribution of the interface-like
phonons to Raman spectra rises when increasing $R/h.$

We have shown that non-adiabatic transitions \emph{drastically enhance the
efficiency of the exciton-phonon interaction in quantum dots}. Another
mechanism, which is responsible for this enhancement, is a separation of the
electron and hole charges owing to different masses of an electron and of a
hole, when a finite potential barrier exists at the quantum-dot interface.
When strengthening confinement, the efficiency of the exciton-phonon
interaction rises.

In view of the results as obtained here, it would be important to
experimentally investigate \emph{multiphonon} Raman spectra in disk-shape
self-assembled quantum dots.

\bigskip

\begin{center}
\textbf{ACKNOWLEDGMENTS}

\bigskip
\end{center}

We are grateful to E. P. Pokatilov and V. N. Gladilin for valuable
discussions. This work has been supported by the BOF NOI (UA-UIA), GOA BOF UA
2000, IUAP, FWO-V. projects G.0287.95, 9.0193.97, G.0274.01N and the W.O.G.
WO.025.99N (Belgium).

\newpage

\begin{quote}
\textbf{Figure captions}
\end{quote}

\bigskip

Fig. 1. One-phonon (upper panel) and two-phonon (lower panel) Raman spectra
for a cylindrical GaAs/AlAs quantum dot with the height $h=5$ nm and with the
lateral-confinement frequency parameter $\hbar\Omega_{0}=3$ meV. Solid curves
show results obtained taking into account effects of non-adiabaticity. Dashed
curves show results of the adiabatic approximation.\newline \emph{Inset}:
measured Raman spectra of a disk-shape self-assembled Ga$_{0.8}$In$_{0.2}$As
quantum dot (from Refs. \cite{3,4}) for different excitation frequencies. For
clearness, different Raman spectra are shifted in the horizontal plane. The
dot-dashed line indicates the position of the frequency shift $\Omega
_{I}-\Omega_{s}=300$ cm$^{-1}$.

\bigskip

Fig. 2. One-phonon (upper panel) and two-phonon (lower panel) Raman spectra
for a cylindrical GaAs/AlAs quantum dot with the height $h=5$ nm and with the
lateral-confinement frequency parameter $\hbar\Omega_{0}=10$ meV. Denotations
are the same as those in Fig. 1.

\bigskip

Fig. 3. One-phonon (upper panel) and two-phonon (lower panel) Raman spectra
for a cylindrical GaAs/AlAs quantum dot with the height $h=2$ nm and with the
lateral-confinement frequency parameter $\hbar\Omega_{0}=3$ meV. Denotations
are the same as those in Fig. 1.

\bigskip

Fig. 4. The parameter $S$ determined as $S\equiv2I_{2}/I_{1}$, where $I_{1}$
and $I_{2}$ are integral intensities of the one-phonon and two-phonon Raman
bands, respectively, as a function of the confinement frequency $\Omega_{0}$.
In the adiabatic approximation, $S$ is just the Huang-Rhys parameter. The
upper group of curves shows results obtained taking into account non-adiabatic
transitions.The lower group of curves corresponds to the adiabatic
approximation. \newline \emph{Inset}: the ratio of the ``non-adiabatic'' and
``adiabatic'' parameters $S^{\left(  \mathrm{NA}\right)  }/S^{\left(
\mathrm{A}\right)  }$ as a function of the confinement frequency $\Omega_{0}$.
\end{document}